\documentclass[aps,twocolumn,pre,floatfix]{revtex4-2}

\usepackage{xcolor,hyperref}
\hypersetup{
   colorlinks,
   linkcolor={blue!50!black},
   citecolor={blue!50!black},
   urlcolor={blue!80!black}
}

\usepackage{epsfig,amsmath}
\usepackage{bm}
\usepackage{soul,ulem}
\usepackage{hyperref}
\usepackage{footmisc}
\usepackage{wrapfig}
\DeclareMathAlphabet{\mathitbf}{OML}{cmm}{b}{it}

\renewcommand{\=}{\!=\!}

\newcommand{\xv}{\mathitbf x}
\newcommand{\Xv}{\mathitbf X}

\newcommand{\calBold}[1]{\mbox{\boldmath${\cal #1}$}}
\newcommand{\dbar}{{\,\mathchar'26\mkern-12mu d}}

\DeclareMathAlphabet\mathbfcal{OMS}{cmsy}{b}{n}
\begin{document}

\title{Large dilatational hyperelasticity of glasses en route to cavitation failure}
\author{Pawandeep Kaur$^{1}$}
\email{Contributed equally}
\author{Noam Ottolenghi$^{1}$}
\email{Contributed equally}
\author{Edan Lerner$^{2}$}
\author{David Richard$^{3}$}
\email{david.richard@espci.fr}
\author{Eran Bouchbinder$^{1}$}
\email{eran.bouchbinder@weizmann.ac.il}
\affiliation{$^{1}$Chemical and Biological Physics Department, Weizmann Institute of Science, Rehovot 7610001, Israel\\
$^{2}$Institute for Theoretical Physics, University of Amsterdam, Science Park 904, 1098 XH Amsterdam, The Netherlands\\
$^{3}$PMMH, CNRS, ESPCI Paris, Universit\'{e} PSL, Sorbonne Universit\'{e}, Universit\'{e} Paris Cit\'{e}, France}

\begin{abstract}
Materials deform elasto-plastically and fail under various loading conditions, typically quantified by the stress triaxiality, which is the ratio between the dilatational (hydrostatic) stress and the deviatoric (shear-like) one. We show that the elasto-plastic deformation of glasses approaching failure qualitatively differ for large and small stress triaxiality levels. Specifically, in the former limit, glasses reveal a strong hyperelastic (nonlinear elastic) response with minute plasticity, largely independently of the quenching rate across the glass transition. Yet, glassy disorder gives rise to significant elastic (reversible) nonaffine deformation, accompanied by the formation of micro-cavities. A small fraction of the latter is irreversible, i.e., survives unloading prior to the onset of failure, and may serve as nucleation sites for failure in the form of large-scale cavitation, involving a topological transition accompanied by the formation of an internal free surface, upon which the glass loses a significant fraction of its load-bearing capacity. These results are contrasted with glass behavior in the limit of vanishing stress triaxiality and their universality across different glass formers is demonstrated. Finally, the implications of our findings for understanding glass deformation and failure under realistic stress conditions are discussed.
\end{abstract}

\maketitle

\section{I\lowercase{ntroduction}}

The ways materials deform and fail depend not only on the magnitude of the force applied to them, but also on its symmetry, i.e., on the tensorial nature of the material stress state. The latter is commonly quantified by the stress triaxiality~\cite{timoshenko1983history}, which measures the relative magnitude of the hydrostatic (dilatational, volumetric) and deviatoric (shear-like, volume-preserving) stress. While the effect of stress triaxiality on the elasto-plastic deformation and failure modes of crystalline, polycrystalline and porous ductile solids has been extensively studied~\cite{benzerga2016ductile,mcclintock1968criterion,rice1969ductile,lemaitre2012course,lemaitre1985continuous,lemaitre1984use,lemaitre1986local,Chaboche1988PartI,Chaboche1988PartII,bonora1997nonlinear,wang1992unified,chandrakanth1995isotropic,la2001effect,gurson1977continuum,tvergaard1984analysis,needleman1984analysis,tvergaard2002two},
characterizing and understanding the corresponding physical effects in non-crystalline, glassy/amorphous solids lag behind (though, note~\cite{an2011atomistic,guan2013cavitation,lei2015notch,meng2024stress}, for example).

Much effort has been devoted to studying the micromechanics and statistical-mechanical properties of deforming glasses in the limit of vanishing stress triaxiality, i.e., under predominantly shear loading. Realistic failure, however, typically takes place under significantly different physical conditions. For example, the stress state ahead of the tip of a quasi-static crack defect in quasi-two-dimensional systems is purely hydrostatic~\cite{lawn,fineberg.99} and high stress triaxiality levels emerge in various three-dimensional loading configurations~\cite{atroshenko2010deformation,hufnagel2013crack,gu2014mechanisms,pan2015origin,pan2017ductile,rong2022molecular,cheng2022atomistic}. Elevated stress triaxiality is indicated by the formation of micro-cavities near the crack tip~\cite{falk1999molecular,murali2011atomic,rycroft2012fracture,singh2016cavitation,shen2021observation,tang2022crack}, as well as by the appearance of dimpled features in the fractography of ductile Bulk Metallic Glasses~\cite{sun2015fracture,richard2023bridging}.

Preliminary steps in closing the aforementioned gap have been recently made~\cite{guan2013cavitation,dattani2022universal,moriel2024elementary}. Specifically, some similarities between the low-temperature elementary processes that mediate structural rearrangements in glasses under large and small stress triaxiality levels have been demonstrated~\cite{bonfanti2019elementary,dattani2022universal,moriel2024elementary}. Earlier work established that under shear loading, structural rearrangements in glasses correspond to saddle-node bifurcations (instabilities) in the potential energy landscape, whose complexity reflects the disordered nature of glasses, formed by quenching liquids through the glass transition. These instabilities, termed shear transformations, are localized in space and are the main carriers of plastic deformation in glasses under shear~\cite{Argon1979,Falk1998}, playing analogous roles to mobile dislocations in crystalline solids~\cite{anderson2017theory}. It has been recently shown that the very same instabilities occur under large stress triaxiality levels (predominantly dilatational deformation), where the symmetry of the loading quantitatively affects the geometry of the unstable modes~\cite{moriel2024elementary}.

Here, we show that in fact the elasto-plastic deformation of glasses approaching failure qualitatively differ for large and small stress triaxiality levels. By employing large-scale computer simulations and an array of theoretical and computational tools, we show that under large stress triaxiality, glasses reveal a strong hyperelastic (nonlinear elastic) response with minute plasticity, largely independently of the quench rate through the glass transition, which strongly affects the state of glassy disorder/stability. Consequently, dilatational elastic softening is accounted for by a first-principles zero-strain nonlinear elastic expansion, which is terminated by a large-scale cavitation instability. The latter is a topological transition upon which an internal free surface is spontaneously formed.

The dilatational deformation includes a significant nonaffine component --- a manifestation of the intrinsically disordered nature of glasses ---, which is predominantly reversible and accompanied by the formation of micro-cavities. A small fraction of the latter is irreversible, i.e., survives unloading prior to the onset of aforementioned large-scale cavitation, and may serve as nucleation sites for the latter, upon which the glass loses a significant fraction of its load-bearing capacity. These results are contrasted with the corresponding behavior of sheared glasses. The generality of our findings, which have important implications for understanding the deformation and failure dynamics of glasses under realistic stress conditions, is demonstrated across various classes of computer glasses.

\vspace{-0.2cm}
\section{M\lowercase{acroscopic shear and dilatational glassy response}}\vspace{-0.2cm}

Obtaining fundamental insight into glassy deformation requires access to small lengthscales that are currently out of reach experimentally. We consequently resort to computer simulations that open a unique window to such scales, yet allowing to simulate rather large systems and to extract macroscopic response quantities. We first study a polydisperse, Lennard-Jones-type (LJ) computer glass~\cite{sticky_spheres_part_1}, whose main advantage is that it allows to obtain glasses that span a wide range of thermal annealing states (quench rates) upon glass formation, including deeply annealed states that feature glass disorder/stability comparable to laboratory glasses~\cite{LB_swap_prx}. All the glass samples we consider feature a vanishingly small initial pressure, representing realistic ambient conditions.

We also employ an athermal quasistatic (AQS) deformation procedure that mimics the limit of low temperatures and strain rates~\cite{lemaitre2004_avalanches}. Its main advantage is that it allows to disentangle the effects of thermal fluctuations and finite deformation rates from the dominant effect of glassy disorder on the deformation, and to obtain --- both theoretically and numerically --- quantitative information about various physical observables. The details of the computer model, glass formation procedure and loading protocols are provided in the Methods and {\color{blue}{\em SI Appendix}}.

\begin{figure}[ht!]
    \centering
    \includegraphics[width=0.5\textwidth]{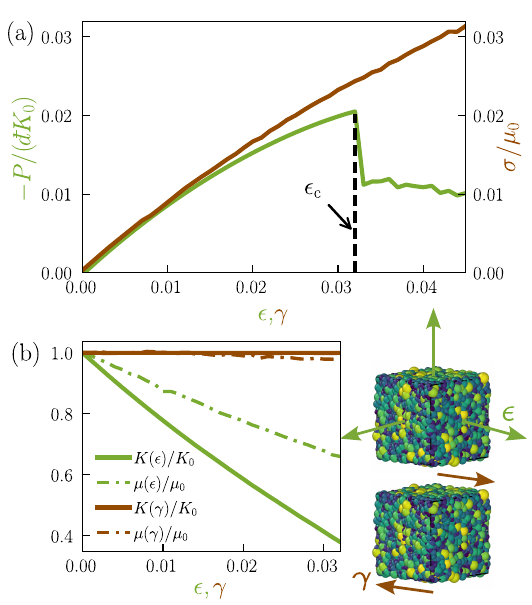}
    \vspace{-0.75cm}
    \caption{(a) Stress-strain curves of the same glass sample ($N\!=\!10^6$ particles) under dilation (green, left $y$-axis) and shear (brown, right $y$-axis), see also visual inset in panel (b). The hydrostatic tension $-P$ (shear stress $\sigma$) is plotted against the dilatational (shear) strain parameter $\epsilon$ ($\gamma$), both normalized such that the dimensionless zero-strain modulus equals unity, where $K_0$ ($\mu_0$) is the bulk (shear) modulus and $\dbar$ is space dimensionality (here $\dbar\!=\!3$). $\epsilon_{\rm c}$ corresponds to the large-scale cavitation strain under dilation (see Fig.~\ref{fig:fig6}). See~{\color{blue}{\em SI Appendix}} for details and definitions. (b) The corresponding strain-dependent bulk and shear moduli (up to the value of $\epsilon_{\rm c}$), see legend.}
    \vspace{-0.5cm}
    \label{fig:fig1}
\end{figure}

We subject the very same glasses to both shear (vanishing stress triaxiality), quantified by a shear strain parameter $\gamma$, and hydrostatic tension/dilation (arbitrarily large stress triaxiality), quantified by a dilatational strain parameter $\epsilon$, and focus first on global quantities. We employ computer glasses composed of $N\=10^4\!-\!10^6$ atoms/particles, depending on the question posed and computational feasibility. In Fig.~\ref{fig:fig1}a, we present normalized stress-strain curves for a quickly quenched glass (i.e., a poorly annealed one, resulting in a highly disordered state) under shear (brown) and dilation (green). Here, and throughout the paper, we use the zero-strain shear (bulk) modulus $\mu_0$ ($K_0$) to normalize the shear stress (hydrostatic tension).

A clear signature of the highly disordered state of the glass sample used is the lack of a distinct yielding transition under shear, i.e., it features a continuous crossover from a linear elastic regime at small $\gamma$ to steady flow at large $\gamma$ (saturating at a flow stress of $\sigma/\mu_0\!\simeq\!0.05$, see~{\color{blue}{\em SI Appendix}}), with no shear-banding. This is in sharp contrast to the corresponding behavior under dilation, which reveals a stronger softening response and a sharp stress drop, corresponding to a large-scale cavitation instability, followed by continuous softening (for the response at yet larger dilatational strains, see~{\color{blue}{\em SI Appendix}}). We hereafter denote the large-scale cavitation strain by $\epsilon_{\rm c}$ and focus on strains smaller or equal to it. Our major goal is to elucidate the origin and physical significance of this difference.

To start characterizing the glass states along the stress-strain loading curve shown in Fig.~\ref{fig:fig1}a, we present in Fig.~\ref{fig:fig1}b the evolution of the {\em strain-dependent linear elastic} constants $\mu$ and $K$ as a function of $\epsilon$ and $\gamma$ (up to a value corresponding to $\epsilon_{\rm c}$), computed from analytic expressions (see~{\color{blue}{\em SI Appendix}}). These should be distinguished from the tangent moduli (that correspond to the local derivatives of each stress with respect to the relevant strain parameter, see~{\color{blue}{\em SI Appendix}}).

The results reveal sharp differences between shear and dilation; while the linear elastic moduli are essentially independent of $\gamma$ under shear, the corresponding quantities feature significant softening under dilation, where the bulk modulus $K(\epsilon)$ softens by $\sim\!60\%$ as $\epsilon_{\rm c}$ is approached and the shear modulus $\mu(\epsilon)$ by about half of it. The results in Fig.~\ref{fig:fig1} may suggest that on the one hand, the dilatational response is ``more brittle'' (likewise, the shear response is ``more ductile'') as it is accompanied by an abrupt and significant stress drop, but at the same time it is also ``less brittle'' as the stress drop is preceded by significant softening.\\

\vspace{-0.2cm}
\section{M\lowercase{icroscopic measures of nonaffinity and structural rearrangements}}\vspace{-0.2cm}

\begin{figure}[ht!]
    \centering
    \includegraphics[width=0.5\textwidth]{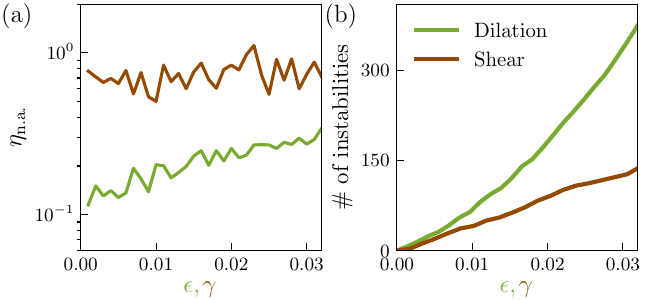}
    \vspace{-0.3cm}
    \caption{(a) The dimensionless measure of nonaffine deformation $\eta_{_{\rm n.a.}}$ vs.~strain under both dilation and shear (as in Fig.~\ref{fig:fig1} but prior to large-scale cavitation and with $N\!=\!10^5$), see text for details and discussion. (b) The corresponding accumulated number of structural rearrangements (instabilities).}
    \vspace{-0.1cm}
    \label{fig:fig2}
\end{figure}

To shed additional light on these somewhat conflicting interpretations, we next consider microscopic observables related to the disordered nature of glasses and their structural evolution under deformation. A distinguishing feature of glassy disorder is nonaffine deformation, i.e., a local deformation that does not follow the globally applied deformation due to disorder-induced mismatch forces~\cite{maloney2006amorphous,athermal_elasticity_2009}. We construct a dimensionless measure of nonaffinity denoted as $\eta_{_{\rm n.a.}}$, which quantifies the fraction of the deformation that cannot be accounted for by a local affine deformation, allowing to compare on equal footing nonaffinity under shear and dilation, see~{\color{blue}{\em SI Appendix}}. In Fig.~\ref{fig:fig2}a, we plot $\eta_{_{\rm n.a.}}$ for a sheared and dilated glass. It is observed that the overall nonaffinity is significantly larger under shear.

In many cases, nonaffinity is related to the structural evolution of the glass due to configurational rearrangements. Therefore, we take advantage of the employed AQS deformation, which allows to exhaustively identify all structural rearrangement/instability events, each one corresponding to a single saddle-node bifurcation. We enumerate all instabilities under shear and dilation, and present in Fig.~\ref{fig:fig2}b their accumulated number as a function of strain. It is observed that the rate (per unit strain) of structural rearrangements is larger under dilation (by a factor somewhat larger than two).

It is thus tempting to hypothesize that the increased rate of structural rearrangements might explain the dramatic excess of softening under dilation observed in Fig.~\ref{fig:fig1}b. Yet, the nature of the structural rearrangements and their magnitude, not just their rate, can play important roles as well. Taken together, the results presented in Fig.~\ref{fig:fig2} do not offer a conclusive physical explanation for the fundamental differences between the glass response to shear and dilation observed in Fig.~\ref{fig:fig1}.

\vspace{-0.2cm}
\section{M\lowercase{acroscopic reversibility and irreversibility revealed by unloading}}\vspace{-0.2cm}

\begin{figure}[ht!]
    \centering
    \includegraphics[width=0.48\textwidth]{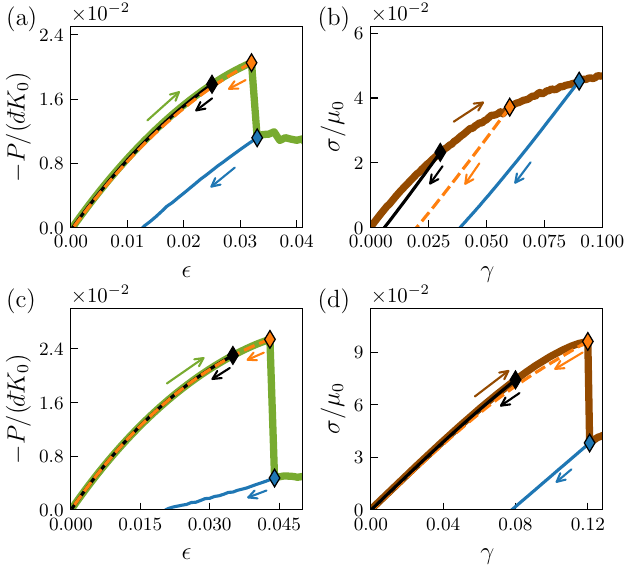}
    \vspace{-0.4cm}
    \caption{(a-b) The same as Fig.~\ref{fig:fig1}a, but with unloading to zero stress added (the unloading points are marked by diamonds and guiding arrows are added). (c-d) The same as panels (a-b), but for a deeply annealed (slowly quenched) glass sample. See text for discussion.}
    \vspace{-0.4cm}
    \label{fig:fig3}
\end{figure}

The above results showed that nonaffinity and structural rearrangements, which are generic features of glassy deformation, are present under both shear and dilation, in line with recent findings~\cite{dattani2022universal,moriel2024elementary}, yet the emerging quantitative differences appear to fall short of accounting for the qualitative differences observed in Fig.~\ref{fig:fig1}. What is missing is a deeper understanding of which of (or the degree to which) the aforementioned deformation processes are reversible or irreversible. Put differently, we need to identify the elastic and plastic components of the deformation under the different stress triaxiality levels. The most definitive way to achieve this is by unloading the glass from different loading levels, as done in Figs.~\ref{fig:fig3}a-b.

The results reveal striking differences between shear and dilation. Figure~\ref{fig:fig3}a shows that unloading dilatational states with $\epsilon\!<\!\epsilon_{\rm c}$ leads to a recovery of the macroscopic initial state, with no significant hysteresis and residual plastic strain. It indicates that minute dissipation is involved in the loading process up to $\epsilon_{\rm c}$, i.e., the corresponding deformation is predominantly elastic/reversible. This means that the dilatational softening observed in Fig.~\ref{fig:fig1}b mainly corresponds to hyperelasticity, i.e., to nonlinear elasticity, not to plasticity. Likewise, the dilatational nonaffinity of Fig.~\ref{fig:fig2}a is expected to be predominantly elastic and the corresponding structural rearrangements quantified in Fig.~\ref{fig:fig2}b to be small and/or reversible upon unloading. Significant hysteresis and residual plastic strains emerge only after the large stress drop occurring at $\epsilon_{\rm c}$, as shown in Fig.~\ref{fig:fig3}a.

The situation is markedly and qualitatively different when the very same glass is unloaded after being sheared, as shown in Fig.~\ref{fig:fig3}b. As expected from a continuous elasto-plastic yielding transition, the unloading curves observed in Fig.~\ref{fig:fig3}b feature a slope approximately corresponding to $\mu_0$, and reveal large hysteresis and residual plastic strains. These are consistent with the absence of elastic softening under shear, as observed in Fig.~\ref{fig:fig1}b, and show that the shear nonaffinity of Fig.~\ref{fig:fig2}a is predominantly plastic and that the corresponding structural rearrangements quantified in Fig.~\ref{fig:fig2}b are predominantly irreversible plastic events.

Taken together, the results in Fig.~\ref{fig:fig3} explain the origin of the fundamental shear/dilation differences of Fig.~\ref{fig:fig1}, and shed basic light on the findings of Fig.~\ref{fig:fig2}. The crux of the differences is that under shear, glass deformation is strongly plastic/irreversible, while under dilation it is predominantly elastic/reversible, revealing large hyperelasticity.

\vspace{-0.2cm}
\section{N\lowercase{onequilibrium thermal history effects}}\vspace{-0.2cm}

The structural rearrangements under dilation quantified in Fig.~\ref{fig:fig2}b do not give rise to significant plastic dissipation and stress relaxation. This implies that they are either small in magnitude or reversible upon unloading or both. Reversible structural rearrangements, i.e., saddle-node bifurcations that are elastically generated during loading, are documented for well annealed glasses deformed under shear~\cite{xu2017strain,jin2018stability,kapteijns2019fast,Richard2020}. We are therefore interested in exploring the differences in the deformation and failure modes of glasses under vastly different stress triaxiality levels also as a function of the glass thermal history.

The glass state analyzed in Fig.~\ref{fig:fig3}a-b was generated through a quick quench, leading to a large degree of disorder. We control the degree of annealing of a glass as it forms, i.e., its effective quench rate, through the `parent temperature' $T_{\rm p}$ from which the equilibrium supercooled liquid is instantaneously brought into a zero-temperature inherent structure~\cite{heuer2008exploring}. The employed computer glass can be very deeply annealed, leading to very stable glass states~\cite{singh2013ultrastable,LB_swap_prx,parmar2020ultrastable}.

In Fig.~\ref{fig:fig3}c-d, we present loading-unloading curves as in Fig.~\ref{fig:fig3}a-b, but for a much deeper supercooled glass state. The response to dilatational deformation, shown in Fig.~\ref{fig:fig3}c, is qualitatively similar to the corresponding dilatational response in Fig.~\ref{fig:fig3}a for a significantly less annealed glass, where the main quantitative difference is the much larger stress drop in the former. The response to shear deformation in Fig.~\ref{fig:fig3}d, however, is dramatically different from the corresponding shear response in Fig.~\ref{fig:fig3}b for a significantly less annealed glass. Notably, the loading-unloading curves in Fig.~\ref{fig:fig3}d (observed earlier, e.g., in~\cite{jin2018stability,kapteijns2019fast}) are qualitatively similar to the dilatational response in Fig.~\ref{fig:fig3}a,c, revealing no macroscopic residual plastic strain prior to the large stress drop.

These results point to yet another major difference between the shear and dilatational response of glasses, indicating that while the dilatational response is predominantly elastic prior to large-scale cavitation, largely independently of the glassy state of disorder (degree of thermal annealing), the macroscopic reversibility-irreversibility properties of the shear response strongly depend on it. The shear and dilatational response in Fig.~\ref{fig:fig3}c-d differ in two additional important respects: first, the physical (and spatial) origin of the large stress drops observed in the two cases is markedly different, i.e., it is large-scale cavitation under dilation (see Fig.~\ref{fig:fig6}a and~{\color{blue}{\em SI Appendix}}), while it is large-scale shear-banding under shear~\cite{varnik2003shear,shi2005strain,Ozawa6656}, having no topological signature. Second, nonlinear elasticity under shear prior to shear-banding is very small, while it is very large under dilation, i.e., $K$ can soften up to $75\%$ over a dilatational strain of $3\!-\!4\%$ (see~{\color{blue}{\em SI Appendix}}, where the deeply annealed counterpart plot of Fig.~\ref{fig:fig1}b is presented).

\vspace{-0.2cm}
\section{M\lowercase{icroscopic irreversibility}}\vspace{-0.2cm}

\begin{figure}[ht!]
    \centering
    \includegraphics[width=0.5\textwidth]{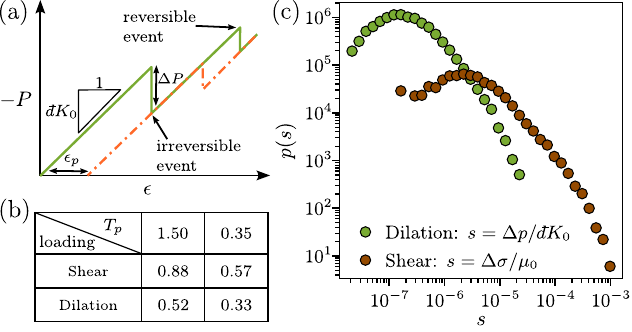}
    \vspace{-0.6cm}
    \caption{(a) A stress-strain curve sketch, showing a loading portion (solid green line) followed by unloading (dashed-dotted orange line). It features a reversible structural rearrangement (event) and an irreversible one, resulting in a residual plastic strain $\epsilon_{\rm p}$. The sketch refers to dilation such that the slope corresponds to $\dbar K_0$ (see Fig.~\ref{fig:fig1}a) and the irreversible stress drop to $\Delta{P}$. (b) A table showing the fraction of irreversible events among the population of the first structural rearrangements in two glass ensembles of $M\!=\!100$ samples each and $N\!=\!10^4$, corresponding to high/low $T_{\rm p}$ values (horizontal direction) and loaded in shear/dilation (vertical direction). See text for discussion. (c) The distribution of dimensionless stress drops $s$ under both dilation and shear (see legend), obtained for $M\!=\!100$ high $T_{\rm p}$ and $N\!=\!10^5$ glasses up to $3\%$ strain.}
    \vspace{-0.3cm}
    \label{fig:fig4}
\end{figure}

In Fig.~\ref{fig:fig3}, we focused on macroscopic reversibility/irreversibility. 
Yet, it was also established above that microscopic glass deformation is accompanied by nonaffinity and structural rearrangements, which can be either irreversible --- and of variable magnitude --- or reversible. Our goal here is to gain insight into {\em microscopic} reversibility/irreversibility at the level of individual structural rearrangements as a function of both stress triaxiality and the glass thermal history. In Fig.~\ref{fig:fig4}a, we illustrate the stress-strain portrait of both reversible and irreversible structural rearrangements/instabilities (`events').

Performing a microscopic reversibility/irreversibility analysis for individual structural instabilities over large strain intervals and system sizes is a daunting computational task. We therefore performed the analysis on the first instability in $N\=10^4$ glass samples generated under the same widely different thermal histories considered in Fig.~\ref{fig:fig3}, quantified by $T_{\rm p}$, and deformed under dilation and shear (loading to the first instability and unloading back to zero stress). In each of these 4 physical conditions, we quantified the fraction of irreversible instabilities of the first instability in $M\=100$ samples, as summarized in Fig.~\ref{fig:fig4}b. It is observed that the fraction of irreversible first instabilities is largest and close to unity for the high $T_{\rm p}$ samples (quick quench, poor annealing) under shear (small stress triaxiality) and that it decreases with both decreasing $T_{\rm p}$ (slower quench, better annealing) and increasing stress triaxiality (here, the dilatational limit). These results indicate that structural rearrangements under large stress triaxiality tend to be more reversible.

Next, we quantify the magnitude of structural rearrangements, whether reversible or irreversible, under shear and dilation. In Fig.~\ref{fig:fig4}c, we present the probability $p(s)$ of dimensionless stress drops $s$ (see legend) during structural rearrangements under shear and dilation for quickly quenched (high $T_{\rm p}$) samples up to $3\%$ strain (smaller than $\epsilon_{\rm c}$). It is observed that stress drops under dilation are more than an order of magnitude smaller than those under shear. Hence, not only structural rearrangements under dilation tend to be more reversible but they are also significantly less intense, outweighing their larger rate observed in Fig.~\ref{fig:fig2}b and consistently with the significantly lower dilatational nonaffinity observed in Fig.~\ref{fig:fig2}a. Overall, these results explain the main observation in Fig.~\ref{fig:fig3} that dilatational deformation up to large-scale cavitation is macroscopically reversible/elastic, largely independently of the glass thermal history. They do not, in themselves, explain its nonlinearity, i.e., the large dilatational hyperelasticity.

\vspace{-0.2cm}
\section{F\lowercase{irst-principles zero-strain nonlinear elastic expansion}}\vspace{-0.2cm}

The large observed dilatational hyperelasticity is unusual. The most common and well-understood origin of hyperelasticity is entropic elasticity (`rubber elasticity') observed in soft solids such as rubber and polymers~\cite{treloar1975physics}. It typically emerges at large strains and its degree, i.e., the fractional change in the elastic modulus (either softening or stiffening) per unit strain, is usually rather mild. Stiffening hyperelasticity is also well-documented in underconstrained (sub-isostatic), disordered biopolymer networks and colloidal gels~\cite{storm2005nonlinear,de2019irreversible,lerner2023scaling,zhang2025strain}, which are qualitatively different from the dense glasses we studied. Our results suggest that dilatational hyperelasticity of glasses can be large, occur under small strains and its origin is energetic, not entropic. That is, they indicate that the observed large hyperelastic softening emerges from the nonlinearity of the interatomic interactions in the absence of significant dilatational irreversibility/plasticity.

This physical picture is translated into a strong theoretical prediction  that the observed large hyperelastic softening under dilation quantitatively follows from the zero-strain ($\epsilon\=0$) nonlinear elastic expansion
\begin{equation}
\frac{-P(\epsilon)}{\dbar K_0}=\epsilon + \tilde{K}_2\,\epsilon^2 + {\cal O}(\epsilon^3) \ .
\label{eq:nonlinear_expansion}
\end{equation}
Here, $K_0$ is the bulk modulus as above, $\dbar$ is the spatial dimension (see~{\color{blue}{\em SI Appendix}}) and $\tilde{K}_2\!\equiv\!K_2/(\dbar K_0)$ is the dimensionless second-order dilatational elastic constant. In the~{\color{blue}{\em SI Appendix}}, we present first-principles expressions for $K_0$ and $K_2$ involving only the particle positions and interaction potential in the $\epsilon\=0$ state. That is, we derive the nonlinear elastic expansion in Eq.~\eqref{eq:nonlinear_expansion} based only on the reference $\epsilon\=0$ state, using a nonlinear response theory, involving no actual deformation~\cite{athermal_elasticity_2009}.
\begin{figure}[ht!]
    \centering
    \includegraphics[width=0.48\textwidth]{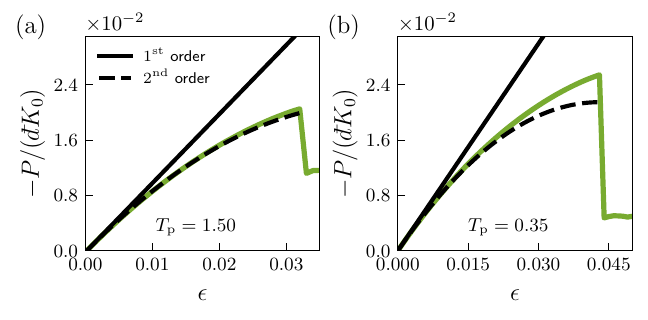}
    \vspace{-0.45cm}
    \caption{(a-b) The same stress-strain curves as in Fig.~\ref{fig:fig3}a,c respectively, where the first-principles zero-strain expansion of Eq.~\eqref{eq:nonlinear_expansion} up to first order (solid line) and second order (dashed line) are superposed. See text for discussion.}
    \vspace{-0.1cm}
    \label{fig:fig5}
\end{figure}

The outcome is presented in Fig.~\ref{fig:fig5}, where the prediction in Eq.~\eqref{eq:nonlinear_expansion} is superposed on the dilatational stress-strain curve of Fig.~\ref{fig:fig3}a,c, for the two widely different thermal histories, revealing excellent quantitative agreement up $\epsilon_{\rm c}$. We emphasize that this great agreement is achieved despite the fact that the loading process is punctuated by discontinuities associated with structural rearrangements/instabilities. The small deviations observed close to $\epsilon_{\rm c}$ correspond to a third-order term in Eq.~\eqref{eq:nonlinear_expansion}, not considered here. These results conclusively demonstrate the hyperelastic nature of dilatational glassy deformation approaching large-scale cavitation and its energetic origin.

\vspace{-0.2cm}
\section{I\lowercase{rreversible micro-cavities and the onset of large-scale cavitation}}\vspace{-0.2cm}

The predominantly hyperelastic dilatational deformation is truncated by a cavitation instability, upon which a large cavity spontaneously forms inside the glass, accompanied by a significant loss of load-bearing capacity. Unlike its shear counterpart, where shear-banding strongly depends on the glass thermal history, this dilatational phenomenology is qualitatively independent of it (quantitative difference do exist, see the different stress drops at different values of $\epsilon_{\rm c}$ in Fig.~\ref{fig:fig5}). A comprehensive analysis and understanding of the irreversible cavitation instability go beyond the scope of this work. Yet, in light of the above results about the predominantly reversible nature of dilatational glassy deformation, it is important to pose the question of the emergence of irreversibility under dilation and the physically-relevant mesoscopic objects that mediate it.

\begin{figure*}[ht!]
    \centering
    \includegraphics[width=\textwidth]{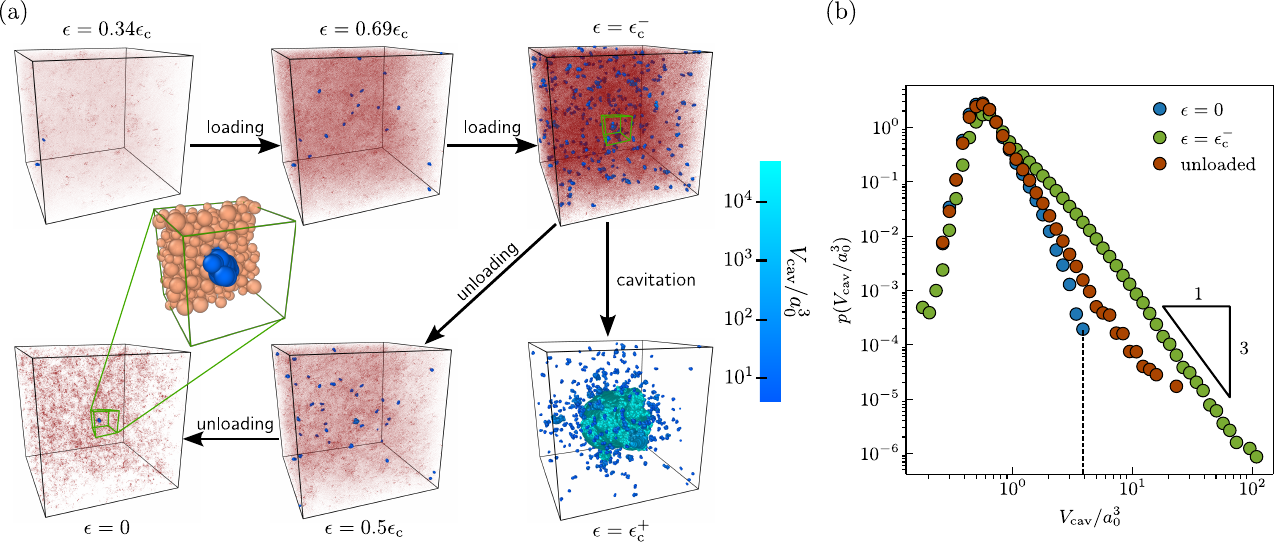}
    \vspace{-0.6cm}
    \caption{(a) Spatial distributions of micro-cavities (labeled by volume $V_{\rm cav}$, see right colorbar) and the magnitude of particle-level nonaffine displacements (in red) along the loading-unloading path of the dilated glass of Fig.~\ref{fig:fig1}a (see strain values $\epsilon$ and labeled arrows). The zoom-in inset shows an irreversible micro-cavity with its surrounding particles, featuring $V_{\rm cav}\!=\!11.3 a_0^3$ with $a_0\!\equiv\![V(\epsilon\!=\!0)/N]^{1/3}$, also marked on the $\epsilon\!=\!\epsilon_{\rm c}^{-}$ snapshot. See text for discussion. (b) $p(V_{\rm cav}/a_0^3$) at $\epsilon\!=\!0$, $\epsilon\!=\!\epsilon_{\rm c}^{-}$ and the unloaded state (see legend), where the tail of the $\epsilon\!=\!\epsilon_{\rm c}^{-}$ distribution follows $\sim\!(V_{\rm cav})^{-3}$. The vertical dashed line denotes the lower threshold for the micro-cavities shown in panel (a).}
    \vspace{-0.5cm}
    \label{fig:fig6}
\end{figure*}

To start addressing this question, it is natural to speculate that the large-scale cavity emerges from smaller-scale cavities that form under dilation and lose stability. To explore this possibility, we developed a micro-cavity detection algorithm (see~{\color{blue}{\em SI Appendix}}) and apply it to a quickly quenched glass undergoing dilatational loading up to large-scale cavitation followed by unloading back to zero stress, and report the results in Fig.~\ref{fig:fig6}.

In Fig.~\ref{fig:fig6}a, we present snapshots of micro-cavities at different pre-cavitation loading strains $\epsilon\!\le\!\epsilon_{\rm c}^{-}$ (top row). It is observed that both the size and number of micro-cavities increase with $\epsilon$, and their spatial distribution is homogeneous. The probability $p(V_{\rm cav})$ of the  micro-cavity volume $V_{\rm cav}$ at $\epsilon\=0$ and $\epsilon\=\epsilon_{\rm c}^{-}$, obtained over an ensemble of $M\=100$ samples, is presented in Fig.~\ref{fig:fig6}b, clearly demonstrating the significant increase in $V_{\rm cav}$ with $\epsilon$. At $\epsilon\=\epsilon_{\rm c}^{-}$, the distribution features a power-law tail $p(V_{\rm cav})\!\sim\!(V_{\rm cav})^{-3}$, apparently independently of the glass thermal history, see~{\color{blue}{\em SI Appendix}}. Next, in Fig.~\ref{fig:fig6}a (bottom row, right), we present the cavities snapshot at $\epsilon\=\epsilon_{\rm c}^{+}$, which reveals a large, quasi-spherical cavity and a cloud of micro-cavities inhomogeneously surrounding it. This cavitation instability is the origin of the large stress drop observed in Figs.~\ref{fig:fig1}a,\ref{fig:fig3}a,\ref{fig:fig5}a.

The observed micro-cavities upon loading are not necessarily irreversible. The physical irreversibility condition is expected to be related to the emergence of a curvature-dependent surface energy. While it is not discussed here, it is clear that once it is met for given micro-cavity, the latter will persist (`survive') upon unloading. A possibly related question concerns the critical nuclei for cavitation, i.e., whether a subset of the micro-cavity population at $\epsilon\=\epsilon_{\rm c}^{-}$ serves as a seed to large-scale cavity formation at $\epsilon\=\epsilon_{\rm c}^{+}$. This is indeed the case, and the relevant micro-cavity is marked on the $\epsilon\=\epsilon_{\rm c}^{-}$ snapshot. Then, snapshots of the micro-cavity population upon unloading are presented (bottom row), demonstrating that the very same micro-cavity survives unloading, i.e., it is irreversible (see zoom-in inset). Finally, $p(V_{\rm cav})$ of unloaded samples is superposed on Fig.~\ref{fig:fig6}b, also revealing a signature of the irreversible micro-cavities at the tail of the distribution.

The scarcity of irreversible micro-cavities is a clear manifestation of the predominantly reversible/elastic nature of the pre-cavitation dilatational deformation of glasses. Yet, these objects are mesoscopic and do not exclude the emergence of atomistic irreversibility. To highlight this point, we superpose on the loading-unloading snapshots in Fig.~\ref{fig:fig6}a the magnitude of the particle-level nonaffine displacements, i.e., the magnitude of the displacement vector relative to the $\epsilon\=0$ configuration, once the global affine deformation is subtracted. It is observed that particle-level nonaffinity exists throughout the sample during the loading process, peaked prior to large-scale cavitation, and that part of it is irreversible, i.e., persists upon unloading.

\vspace{-0.2cm}
\section{U\lowercase{niversality across different glass formers}}\vspace{-0.2cm}

\begin{figure}[ht!]
    \centering
    \includegraphics[width=0.48\textwidth]{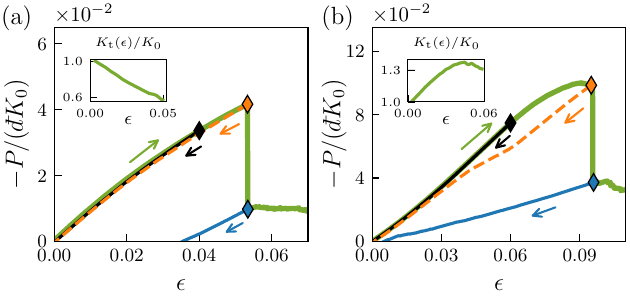}
    \vspace{-0.2cm}
    \caption{(a) The dilation loading-unloading (up to $\epsilon_{\rm c}$) stress-strain curve for a computer CuZr metallic glass, see~{\color{blue}{\em SI Appendix}}. (inset) The loading tangent modulus $K_{\rm t}(\epsilon\!\le\!\epsilon_{\rm c})$. (b) The same as panel (a), but for a computer model of silica glass, see~{\color{blue}{\em SI Appendix}}. (inset) $K_{\rm t}(\epsilon\!\le\!0.06\!<\!\epsilon_{\rm c})$. See text for discussion.}
    \vspace{-0.45cm}
    \label{fig:fig7}
\end{figure}

The results obtained for LJ glasses are expected to be representative of a broader class of glass formers. To explicitly demonstrate this, we studied a CuZr metallic glass using an EAM potential~\cite{mendelev2019development} and a silica glass using the SHIK potential~\cite{sundararaman2020new}, see~{\color{blue}{\em SI Appendix}}. The resulting dilation loading-unloading curves are presented in Fig.~\ref{fig:fig7}a-b, following the same format of Fig.~\ref{fig:fig3}a. Macroscopic reversibility up to large-scale cavitation is observed in these two very different glass formers, significantly supporting the universality of our findings. Note that the large elastic hysteresis loop for silica is characterized by a two-step unloading process: an initial stage in which micro-cavities shrink with minute microscopic nonaffinity, followed by an abrupt increase in nonaffine motions of the glassy network.

The inset in Fig.~\ref{fig:fig7}a shows the loading tangent bulk modulus $K_{\rm t}(\epsilon)$ up to cavitation of the CuZr glass, revealing a strong hyperelastic softening that is quantitatively similar to the corresponding behavior of the LJ glass in Fig.~\ref{fig:fig1}. The corresponding $K_{\rm t}(\epsilon)$ of the silica glass is shown in the inset of Fig.~\ref{fig:fig7}b for small strains, below $\epsilon_{\rm c}$. We focus on small strains since the hyperelastic response in this range reveals stiffening, not softening, highlighting that the degree of hyperelasticity and even its sign may depend on the class of glass formers. The subsequent softening observed in Fig.~\ref{fig:fig7}b at larger strains approaching cavitation, as well as the corresponding shear loading-unloading curves, are presented and discussed in the~{\color{blue}{\em SI Appendix}}. Note that our dilatational silica results might be related to the experimental and simulational uniaxial tension results of~\cite{gupta2005intrinsic,zhang2022origin}, though the basic question of the reversibility/irreversibility of the pre-failure deformation has not been addressed therein.

\vspace{-0.2cm}
\section{S\lowercase{ummary and Outlook}}\vspace{-0.2cm}

We demonstrated that the deformation and failure modes of glasses strongly depend on the stress triaxiality. Specifically, we showed that glasses under dilation, i.e., in the limit of large stress triaxiality, are predominantly elastic up to a large-scale cavitation instability --- typically featuring significant hyperelasticity --- largely independently of the glass thermal history. Dilated glasses feature nonaffinity and structural rearrangements related to their disordered nature, but these are either reversible (elastic) and/or of small magnitude, making little contribution to macroscopic plasticity. They also feature micro-cavities, a small fraction of which serve as critical nuclei for cavitation and may survive unloading. These findings are contrasted with the corresponding glass response under shear, i.e., in the limit of vanishing stress triaxiality.

Our findings are directly relevant for realistic deformation and failure of glasses, which generically take place under finite stress triaxiality levels, in many cases relatively high ones~\cite{hufnagel2013crack,gu2014mechanisms}, and involve material decohesion. They can be tested in experiments that are specifically designed to achieve high stress triaxiality levels. Our results also give rise to a plethora of questions and future investigation directions, briefly mentioned here. First, as already alluded to above, one needs to develop a better understanding of the physical conditions for the formation of irreversible micro-cavities and of the instability criterion associated with large-scale cavitation. Such a criterion may also shed light on the quantitatively different behavior revealed in silica. Moreover, one needs to characterize the spatiotemporal evolution of the cavity, which requires going beyond the AQS limit, and to understand its size selection.

The natural next step would be to consider finite stress triaxiality levels, which are expected to give rise to stronger couplings between shear- and dilation-mediated deformation processes (e.g., there exists evidence that shear softening can facilitate the formation of micro-cavities~\cite{maass2015long,luo2015tensile,richard2023bridging,meng2024stress}), though the roles of dilatational hyperelasticity --- discussed in this work --- in these processes are entirely unexplored. The emerging physical insights should be incorporated into various coarse-grained theoretical efforts, such as elasto-plastic models~\cite{homer2009mesoscale,kondori2016discrete,nicolas2018deformation} and internal-variable based continuum constitutive relations~\cite{henann2008constitutive,Falk2011,rycroft2012fracture}.

\acknowledgements

E.B.~acknowledges support from the Israel Science Foundation (ISF Grant No.~403/24) and the Harold Perlman Family.\\


\appendix




\setcounter{figure}{0}
\renewcommand{\thefigure}{A\arabic{figure}}

\vspace{1.5cm}
\hspace{-0.3cm}{{\color{blue}{\Large{\em SI Appendix}}}}
\vspace{0.5cm}

The goal of this {\color{blue}{\em SI Appendix}} is to provide additional details regarding the tools and methods employed in the manuscript and to offer some additional supporting results.

\vspace{-0.4cm}
\section{I\lowercase{nteratomic potentials and glass formation protocols}}
\label{sec:potentials}

The first system investigated in the manuscript is composed of $N$ polydisperse spheres interacting via a modified Lennard-Jones (LJ) potential~\cite{sticky_spheres_part_1}. The pairwise interaction reads
\begin{equation}
    \varphi(r_{ij}) \!=\!
\left\{
\begin{array}{cc}
\!\!4\varepsilon \bigg[ \big(\frac{\lambda_{ij}}{r_{ij}}\big)^{\!12} - \big(\frac{\lambda_{ij}}{r_{ij}}\big) ^{\!6} \bigg],
     &  \frac{r_{ij}}{\lambda_{ij}} <  x_{\mbox{\tiny min}}  \\\\
\!\!\varepsilon \bigg[a\big(\frac{\lambda_{ij}}{r_{ij}}\big)^{\!12}\!\!-\!b\big(\frac{\lambda_{ij}}{r_{ij}}\big)^{\!6}\nonumber\\
\qquad\qquad\quad+\sum\limits_{\ell=0}^{3}  c_{\mbox{\tiny $2\ell$}} \big(\frac{r_{ij}}{\lambda_{ij}}\big)^{2\ell} \bigg] , & \,x_{\mbox{\tiny min}}\!\le\! \frac{r_{ij}}{\lambda_{ij}}\!<\! x_{\rm c}\\\\
0\,,  &  \frac{r_{ij}}{\lambda_{ij}} \ge x_{\rm c}
\end{array}
\right.,
 \label{eq:potential}
\end{equation}
where $\varepsilon$ is a microscopic energy scale, $x_{\mbox{\tiny min}}$ and $x_{\rm c}$ are the (dimensionless) locations of the minimum of the Lennard-Jones potential and modified cutoff, respectively. $\lambda_{ij}$ are the pairwise length parameters, defined as
\begin{equation}
    \lambda_{ij} = \frac{1}{2}\left(\lambda_i +\lambda_j\right)\left(1-n_a\left|\lambda_i - \lambda_j\right|\right)\ ,
\end{equation}
where $n_a=0.1$ is a non-additivity parameter, and $\lambda_i$ are the particle size parameters drawn from the distribution $p(\lambda)\!\sim\!\lambda^{-3}$, between $\lambda_{\rm min}\=1.0\lambdabar$ and $\lambda_{\rm max}\=2.2\lambdabar$, where $\lambdabar$ is the simulation unit of length. We express the dimensionless cutoff $x_{\rm c}$ in terms of $x_{\mbox{\tiny min}}\!=\!2^{1/6}$, for simplicity, by defining $r_{\rm c}\!\equiv\!x_{\rm c}/x_{\mbox{\tiny min}}$.

In this work, $r_{\rm c}$ is fixed and is equal to $1.5$. The coefficients $a$, $b$ and $\{c_{2\ell}\}$ are chosen such that the attractive and repulsive parts of $\varphi$ and its first two derivatives are continuous at $x_{\mbox{\tiny min}}$ and $x_{\rm c}$. The system is first equilibrated at a parent temperature $T_{\rm p}$ using a SWAP Monte Carlo simulation in the NVT ensemble \cite{LB_swap_prx}. Subsequently, inherent structures are obtained via energy minimization. The number density $\rho\=N/V_0$, where $V_0$ is the initial volume, is set such that the ratio between the pressure $P$ and bulk modulus $K_0$ of the resulting inherent structures is approximately zero. We consider three system sizes of $N\=10^4, 10^5, 10^6$ particles. Throughout the manuscript, lengths are expressed in dimensionless units using the microscopic scale $a_0\=\rho^{-1/3}$. We have considered two parent temperatures, namely a high one $T_{\rm p}\=1.50$ and a low one $T_{\rm p}\=0.35$, corresponding to a fast and a slow quench, and have generated 100 independent samples for each state point.

We further studied two additional realistic glass formers, namely a ductile bulk metallic glass (BMG) and a fragile silica glass. The BMG is modeled as an amorphous equiatomic Cu$_{50}$Zr$_{50}$ binary alloy described by an embedded-atom method (EAM) potential \cite{mendelev2019development}. The silica glass corresponds to a SiO$_2$ network modeled using the SHIK potential \cite{sundararaman2020new}. This force field is known to accurately reproduce the density and elastic constants of experimental silica glasses \cite{zhang2020critical}.

For both systems, we first equilibrate the melt at a high temperature $T_{\rm melt}$ in the NPT ensemble at zero external pressure ($P\=0$ Pa). The equilibrated liquids are subsequently quenched to $T\=50\,\mathrm{K}$ for CuZr and $T\=300\,\mathrm{K}$ for SiO$_2$, well below the corresponding glass transition temperatures, at constant pressure using a cooling rate $\dot{T}$. The system is then further annealed for $100\,\mathrm{ps}$. The final configurations are then relaxed via energy minimization to obtain the corresponding inherent structures.

For CuZr, the system is composed of $N\=108$K particles and we employ $T_{\rm melt}\=1500\,\mathrm{K}$ and a quench rate $\dot{T}\=10^{14}\,\mathrm{K/s}$. For silica, we use $N\=90$K, $T_{\rm melt}\=3600\,\mathrm{K}$, and $\dot{T}\=2.5^{11}\,\mathrm{K/s}$. All molecular dynamics simulations are performed using the massively parallel LAMMPS package~\cite{plimpton1995fast}. The integration time step for the equations of motion is set to $1\,\mathrm{fs}$.

\vspace{-0.3cm}
\section{D\lowercase{eformation protocols}}
\label{sec:deformation}

All mechanical tests in this work are performed in the zero-temperature and zero strain-rate limits, i.e., under athermal quasi-static (AQS) conditions. Any deformed state of a glass is obtained by a global affine transformation described by
$\xv\!=\! \bm{F}\cdot\Xv$ and applied to a cubic system of initial linear size $L_0$, such that $V_0\=L_0^3$. Here, $\Xv$ is the coordinate describing the reference, undeformed glass, and $\xv$ describes the deformed one. Note that since $\bm{F}$ is independent of $\Xv$, we also have $d\xv\!=\! \bm{F}\cdot d\Xv$, which is the conventional definition of the deformation gradient tensor that relates vectorial line elements in the undeformed and deformed configurations.

We realize the two stress triaxiality limits employed in the manuscript by considering the following homogeneous deformation gradient tensors. First, the limit of vanishing stress triaxiality is obtained using
\begin{equation}
\bm{F}_{\rm s}(\gamma) = \left( \begin{array}{ccc}
1&\gamma&0\\
0&1&0\\
0&0&1
\end{array}\right)\ ,
\label{eq:shear}
\end{equation}
where `s' denotes `shear' and $\gamma$ is a parameter that quantifies the amplitude of the applied shear strain. The conjugated stress is the shear stress of magnitude $\sigma$. Second, the limit of large stress triaxiality is obtained using
\begin{equation}
\bm{F}_{\rm d}(\epsilon) = \left(\begin{array}{ccc}e^\epsilon&0&0\\0&e^\epsilon&0\\0&0&e^\epsilon\end{array}\right)\,,
\label{eq:dilation}
\end{equation}
which is parameterized by the dilatational strain amplitude $\epsilon$, corresponding to the logarithmic (Hencky) strain, and `d' denotes `dilation'. The conjugated stress is the hydrostatic tension $-P$ ($P\!<\!0$ is the hydrostatic pressure).

The shear deformation process corresponding to Eq.~\eqref{eq:shear} is volume preserving, while the dilatational one of Eq.~\eqref{eq:dilation} is shape preserving, but varies the volume of the glass according to $V\=e^{\epsilon\dbar}V_0$, where $\dbar$ is space dimensionality (in our case $\dbar\=3)$. Consequently, we have $\epsilon\=\tfrac{1}{\dbar}\log\!\big(V/V_0\big)$. Upon linearization relative to the zero-strain ($\epsilon\=0$) state, we obtain $\epsilon\!\simeq\!\tfrac{1}{\dbar}\big(V/V_0-1\big)\=\varepsilon_{_{\rm V}}/\dbar$, where $\varepsilon_{_{\rm V}}\!\equiv\! V/V_0-1$ is the linearized/infinitesimal volumetric strain. Since the bulk modulus $K_0\!\equiv\!K(\epsilon\=0)$ is given as $K_0\=-V_0\partial{P}/\partial{V}\=-\partial{P}/\partial\varepsilon_{_{\rm V}}$, whenever we want to present (in the manuscript and here) the dimensionless shear and dilatational stresses on equal footing, we employ $-P(\epsilon)/(\dbar K_0)$ as the counterpart of $\sigma(\gamma)/\mu_0$, where $\mu_0\!\equiv\mu(\gamma\=0)$ is the shear modulus.

Using $\gamma$ and $\epsilon$, the elastic moduli of a system can be computed analytically from the system's potential energy $U$ at any strain according to
{\small \begin{align}
    \mu= & \frac{1}{V}\left(\frac{\partial^2U}{\partial\gamma^2} - \frac{\partial^2U}{\partial\gamma\partial\xv}\cdot\left(\frac{\partial^2U}{\partial\xv\partial\xv}\right)^{-1}\cdot\frac{\partial^2U}{\partial\xv\partial\gamma}\right) \label{eq:mu} \ ,\\
    K=&\frac{1}{V\dbar^2}\left(\frac{\partial^2 U}{\partial \epsilon^2} - \dbar\frac{\partial U}{\partial \epsilon} - \frac{\partial^2U}{\partial\epsilon\partial\xv}\cdot\left(\frac{\partial^2U}{\partial\xv\partial\xv}\right)^{-1}\cdot\frac{\partial^2U}{\partial\xv\partial\epsilon}\right) \ . \label{eq:K}
\end{align}}

At each deformation step, the system is first subjected to an affine transformation according to Eq.~\eqref{eq:shear} or Eq.~\eqref{eq:dilation}, employing a small shear strain increment $\delta\gamma$ or $\delta\epsilon$ for volumetric deformations. Following each affine deformation, the system is relaxed back to mechanical equilibrium via energy minimization, thereby removing the nonaffine forces induced by the imposed strain increment. We use a strain step $\delta\gamma\=10^{-4}$ for all glass ensembles, except for the largest LJ system, for which $\delta\gamma\=10^{-3}$ is employed to reduce computational cost. The same protocol is applied during unloading, where the strain increment is reversed, $\delta\gamma\!\to\!-\delta\gamma$, and the procedure is continued until either $-P/(\dbar K_0)$ or $\sigma/\mu_0$ approximately vanishes.

\begin{figure}[h]
    \centering
    \includegraphics[width=\linewidth]{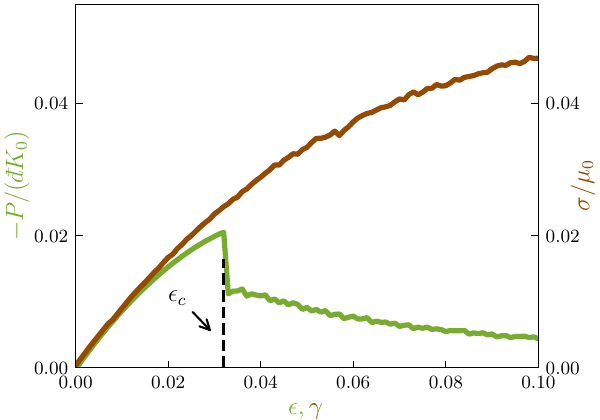}
    \caption{The same stress-strain curves as in Fig.~1 in the manuscript, just going to a larger strain value.}
    \label{fig:stress-strain}
\end{figure}

\begin{figure}[h]
    \centering
    \includegraphics[width=\linewidth]{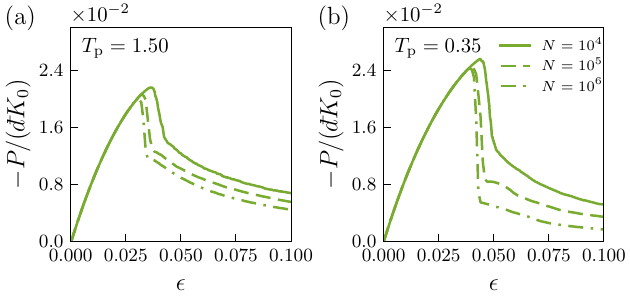}
    \caption{$-P(\epsilon)/(\dbar K_0)$ (averaged over $100$ samples) for (a) poorly-annealed ($T_{\rm p}\!=\!1.50$) and (b) deeply-annealed ($T_{\rm p}\!=\!0.35$) LJ glasses for three system sizes: $N\!=\!10^4, 10^5, 10^6$ (see legend in panel (b)).}
    \label{fig:size}
\end{figure}

\begin{figure}[h]
    \centering
    \includegraphics[width=0.8\linewidth]{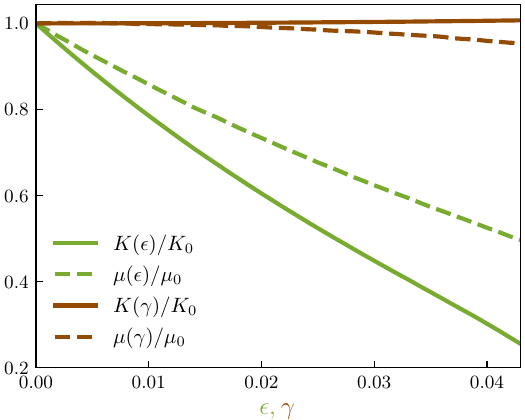}
    \caption{The same as Fig.~1b in the manuscript, but for a deeply-annealed ($T_{\rm p}\!=\!0.35$) LJ glass sample.}
    \label{fig:lowTp_modulus}
\end{figure}

Figure~\ref{fig:stress-strain} shows the stress-strain curve for the same $N\=10^6$ glass sample as shown in Fig.~1 in the manuscript, but for larger strain values up to $\epsilon\, (\text{or}\: \gamma)\!=\!0.1$. The shear stress $\sigma/\mu_0$ saturates at $\sigma\!\simeq\!0.05$ beyond $\gamma\=0.1$, a feature of highly disordered glasses. Figure~\ref{fig:size} shows the hydrostatic stress-strain curve for poorly-annealed ($T_{\rm p}\=1.50$) as well as deeply annealed ($T_{\rm p}\=0.35$) LJ glass ensemble, where each curve is averaged over $100$ samples. We emphasize that in all other stress-strain curves in this work, single samples are used (i.e., no ensemble averages). For both material preparation histories, the hydrostatic stress in the post-yield state, $-P(\epsilon_{\rm c}^{+})$, decreases with increasing system size $N$. This decrease is more pronounced for the well-annealed state. Figure~\ref{fig:lowTp_modulus} shows the strain-dependent shear and bulk moduli under dilation and shear for a deeply annealed glass sample. While the moduli do not vary much under shear $\gamma$, the bulk modulus $K(\epsilon)$ softens by about $75\%$ and $\mu(\epsilon)$ by $\simeq\!50\%$ as $\epsilon_{\rm c}$ is approached.

\vspace{-0.7cm}
\section{A \lowercase{dimensionless nonaffinity measure}}
\label{sec:nonaffinity}

Here, we construct a dimensionless measure of nonaffinity $\eta_{\rm n.a.}$, which quantifies the fraction of the deformation that cannot be described by a local affine deformation. For that purpose, we first introduce a measure of the total (affine and nonaffine) deformation between two glass configurations. It is then used to define a dimensionless measure of nonaffine deformation i.e., the deviation of the total glassy deformation from the one described by a local best affine deformation gradient tensor $\bm{F}_{\rm b.a.}$ (defined below).

To quantify the magnitude of the total deformation, recall the definition of the (rotationally-invariant, metric) Green-Lagrange strain tensor~\cite{holzapfel2002nonlinear}. Consider infinitesimal line elements of size $dl$ and $dl^{\prime}$ in the reference and deformed configurations respectively, and compute the change in their length squared
\begin{multline}
    (dl^{\prime})^2-(dl)^2 = dx_\alpha dx_\alpha-dX_\alpha dX_\alpha =\\
    F_{\alpha\beta}dX_\beta F_{\alpha\gamma}dX_\gamma-dX_\beta\delta_{\beta\gamma}dX_\gamma =\\
    2dX_\beta\bigg[ \frac{1}{2}(F_{\alpha\beta}F_{\alpha\gamma}-\delta_{\beta\gamma})\bigg]dX_\gamma=\\
    2dX_\beta\bigg[\frac{1}{2}(F_{\beta\alpha}^{T}F_{\alpha\gamma}-\delta_{\beta\gamma}) \bigg]dX_\gamma\equiv 2dX_\beta E_{\beta\gamma}dX_\gamma\ ,
\end{multline}
where indices run over spatial dimensions and $\bm{E}\=\frac{1}{2}(\bm{F}^{\rm T}\bm{F}\!-\!1)$ is the Green-Lagrange strain tensor. Similarly, a corresponding scalar strain measure would be $(|d\xv|^2-|d\Xv|^2)/(2|d\Xv|^2)$, which inspires us to define the magnitude of total deformation between two glass configurations, at a point $\Xv^{(i)}$ in space where a particle resides, as
\begin{equation}
    \overline{\big|E^{(i)}|} = \frac{1}{n}\sum_{j=1}^{n} \frac{\Big||d\xv^{(ij)}|^2-|d\Xv^{(ij)}|^2\Big|}{2|d\Xv^{(ij)}|^2} \ .
    \label{Eq:total}
\end{equation}
Here, $d\Xv^{(ij)}$ and $d\xv^{(ij)}$ denote distance between pair of particles in the reference and deformed configurations, respectively, and $n$ is the number of pairs within an interaction distance of twice the typical interparticle distance from the reference particle $i$, i.e., $|d{\bm X}^{(ij)}|\!\le\!2a_0$. That is, $\overline{\big|E^{(i)}|}$ is the average magnitude of deformation in the local environment of any glass particle $i$.

To define the affine deformation, we follow the widely-used $D^2_{\rm min}$ analysis~\cite{Falk1998} and consider the objective function
\begin{equation}
    D^2(\Xv^{(i)}) = \frac{1}{n} \sum_{j=1}^{n}| d\xv^{(ij)}-\bm{F}(\Xv^{(i)})\,d\Xv^{(ij)}|^2 \ ,
    \label{Eq:d2min}
\end{equation}
at any point $\Xv^{(i)}$, using the same pairs of particles as in Eq.~\eqref{Eq:total}. Note that $\bm{F}$ in Eq.~\eqref{Eq:d2min} is consistent with the definition of the deformation gradient tensor in Sect.~\ref{sec:deformation} (see the paragraph above Eq.~\eqref{eq:shear}).

We are interested in $\bm{F}$ that minimizes $D^2(\Xv^{(i)})$ per $\Xv^{(i)}$ and hereafter refer to it as best-affine local deformation gradient tensor $\bm{F}_{\rm b.a}$. The minimal value of $D^2(X^{(i)})$ is commonly denoted as $D^2_{\rm min}(\Xv^{(i)})$~\cite{Falk1998}. Next, we consider the best-affine Green-Lagrange strain tensor $\bm{E}_{\rm b.a}=\frac{1}{2}(\bm{F}_{\rm b.a.}^{\rm T}\bm{F}_{\rm b.a.}\!-\!1)$ and define magnitude of the best-affine deformation as
\begin{equation}
 \overline{|E_{\rm b.a.}^{(i)}|} = \frac{1}{3}\left(|E_{\rm b.a.}^{1,(i)}| + |E_{\rm b.a.}^{2,(i)}| + |E_{\rm b.a.}^{3,(i)}|\right) \ ,
 \label{Eq:affine_deformation}
\end{equation}
which is the average over its eigenvalues $E_{\rm b.a.}^{1},E_{\rm b.a.}^{2},E_{\rm b.a.}^{3}$.

We expect the magnitude of affine deformation in Eq.~\eqref{Eq:affine_deformation} to identify with the one in Eq.~\eqref{Eq:total} in regions of space where the deformation is affine i.e., described by differential operators and proper
strain measures, and to deviate from it in regions where glassy nonaffinity is pronounced. Consequently, we define the following measure of glassy nonaffinity at any point $\Xv^{(i)}$ in space as
\begin{equation}
 \eta_{\rm n.a.}^{(i)} = \Big|\overline{\big|E^{(i)}|}/\overline{|E_{\rm b.a.}^{(i)}|}-1 \Big| \ .
 \label{Eq:non_affine}
\end{equation}
This measure, which unlike $D^2_{\rm min}(\Xv^{(i)})$ itself is dimensionless, is expected to vanish in regions where the deformation is affine and increases in regions where the nonaffinity attains large values.

To test and validate the new dimensionless measure of glassy nonaffinity in Eq.~\eqref{Eq:non_affine}, we need a benchmark glassy deformation field of known properties. A most natural choice would be related to localized structural rearrangements corresponding to saddle-node bifurcations/instabilities, as discussed in the manuscript. Specifically, we consider the displacement field associated with such an instability, defined as the difference between $\Xv$ and $\xv$ corresponding to a configuration just before the instability and a one just after it, respectively. The merit of such a displacement field in the present context is that it is well established that it is highly nonaffine inside its disordered core of size $\xi_{\rm g}$, whose center is located at $r\=0$ in a spherical coordinate system $(r,\theta,\phi)$, and it decays algebraically as $1/r^3$ away from the core, following a dipolar elastic response~\cite{Eshelby1957,Eshelby1959,moriel2024elementary}.

\begin{figure}[h]
    \centering
    \includegraphics[width=0.9\linewidth]{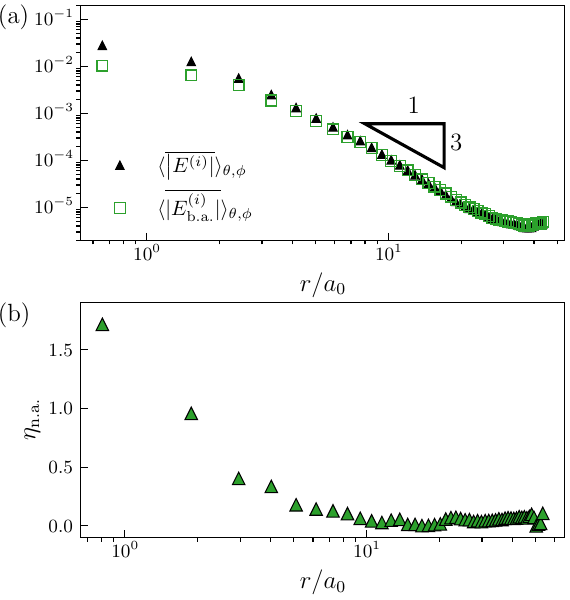}
    \caption{(a) $\langle\overline{\big|E^{(i)}|}\rangle_{\theta,\phi}$ vs.~$r$ (black filled triangles), i.e., the measure of total deformation of Eq.~\eqref{Eq:total} once averaged over the angles $\theta$ and $\phi$, for a displacement field associated with a structural rearrangement under dilation (see text for details). Here, $r\!=\!0$ is the center of the core of the displacement field and the $1/r^3$ away from the core is highlighted in the power-law triangle, see text for discussion. $\langle\overline{|E_{\rm b.a.}^{(i)}|}\rangle_{\theta,\phi}$ vs.~$r$ (green empty squares), i.e., the measure of best-affine deformation of Eq.~\eqref{Eq:affine_deformation}, is superposed. See text for discussion. (b) The corresponding dimensionless nonaffinity measure of Eq.~\eqref{Eq:non_affine}, $\eta_{\rm n.a.}(r)$. See text for discussion.}
    \label{fig:nonaffinity_mode}
\end{figure}

We generated such a displacement field under dilation (see also Sect.~\ref{sec:events}) and plotted in Fig.~\ref{fig:nonaffinity_mode}a the measure of total deformation of Eq.~\eqref{Eq:total} against $r$, i.e., once averaged over the angles $\theta$ and $\phi$. It is observed that, as expected, beyond a core size $\xi_{\rm g}$ of a few $a_0$, $\langle\overline{\big|E^{(i)}|}\rangle_{\theta,\phi}$ decays as $1/r^3$ (see power-law triangle). We then superimpose the measure of best-affine deformation of Eq.~\eqref{Eq:affine_deformation} also against $r$. It is observed that $\langle\overline{|E_{\rm b.a.}^{(i)}|}\rangle_{\theta,\phi}$ deviates from $\langle\overline{\big|E^{(i)}|}\rangle_{\theta,\phi}$ inside the core, but appears to agree with it away from it. Indeed, when plotting in Fig.~\ref{fig:nonaffinity_mode}b the above-developed dimensionless nonaffinity measure of Eq.~\eqref{Eq:non_affine}, $\eta_{\rm n.a.}(r)$, it is observed that the latter is of order unity (corresponding to strong nonaffinity) inside the disordered core of the displacement field associated with the structural rearrangement and essentially vanishes away from it, where the regular elastic $1/r^3$ behavior dominates.

This example demonstrates and supports the validity of the dimensionless nonaffinity measure $\eta_{\rm n.a.}$ of Eq.~\eqref{Eq:non_affine}. Additional examples, applications and more extended analyses of it will be presented elsewhere. An important advantage of $\eta_{\rm n.a.}$ is that it is dimensionless, i.e., it is a strain-like measure of nonaffine glassy deformation, and as such the physical meaning of its values is clear and it can be used to compare nonaffinity in different physical situations. Specifically, in Fig.~2a in the manuscript, it is used to compare on equal footing glassy nonaffinity under shear and dilation. Therein, we present the system-averaged $\eta_{\rm n.a.}$ (i.e., averaged over all $N$ particles) between consecutive configurations separated by $\delta\gamma\,(\text{or} \; \delta \epsilon)\=0.001$.

\vspace{-0.35cm}
\section{D\lowercase{etection of structural rearrangements/instability events}}
\label{sec:events}

Here, we describe the algorithm used to detect structural rearrangements (instability events) along an AQS deformation process. The algorithm closely follows the one described in Sect.~S-2 of the Supplementary Materials file of~\cite{moriel2024elementary}. To identify structural arrangements, we define a measure of their presence in a single deformation step. We denote the nonaffine displacement of each particle, induced by the deformation step of size $\delta\epsilon$ starting at strain $\epsilon$ as ${\bm \Delta}(\epsilon, \delta\epsilon)\=\mathcal{T}_{\rm n.a.} [\xv(\epsilon+\delta\epsilon)\!-\!\xv(\epsilon)]$, where the operator $\mathcal{T}_{\rm n.a.}$ removes the affine part of the deformation. Then, we define the measure for the presence of structural rearrangements in a deformation step as
\begin{equation}
    Q(\epsilon, \delta\epsilon) = \frac{\max\left\Vert{\bm \Delta}(\epsilon, \delta\epsilon)\right\Vert}{\delta\epsilon}\ ,
\end{equation}
where $\max\left\Vert \cdot \right\Vert$ denotes the largest magnitude among all particles. In the absence of rearrangements, nonaffine displacements scale with $\delta\epsilon$, resulting in $Q\={\cal O}(1)$, while a structural rearrangement induces a characteristic nonaffine displacement independently of the strain step, resulting in $Q\={\cal O}({\delta\epsilon}^{-1})$. This property allows us to identify the presence of instability events within a strain step through a large value of $Q$, as well as to identify their precise location by a sharp drop in $Q$. Note that while the notation of $\epsilon$ as strain is used here, this measure is valid for any deformation process, specifically for small and large stress triaxiality levels extensively discussed in the manuscript.

\begin{figure}[h]
    \centering
    \includegraphics[width=\linewidth]{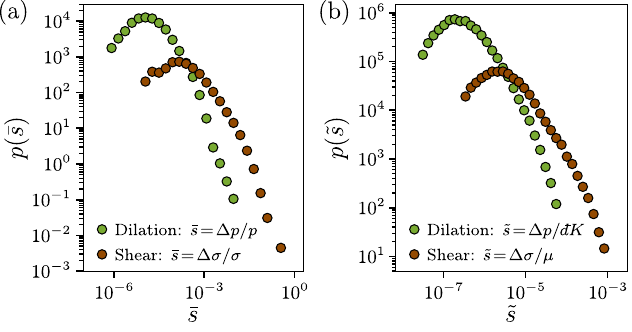}
    \caption{The same structural rearrangement stress drops dataset (under both shear, in brown, and dilation, in green) used in Fig.~4c in the manuscript, but with two alternative nondimensionalization choices, see legends in panels (a) and (b), and the text. See text for the discussion of the results.}
    \label{fig:alternative-drop-dists}
\end{figure}

\begin{figure}[h]
    \centering
     \includegraphics[width=0.9\linewidth]{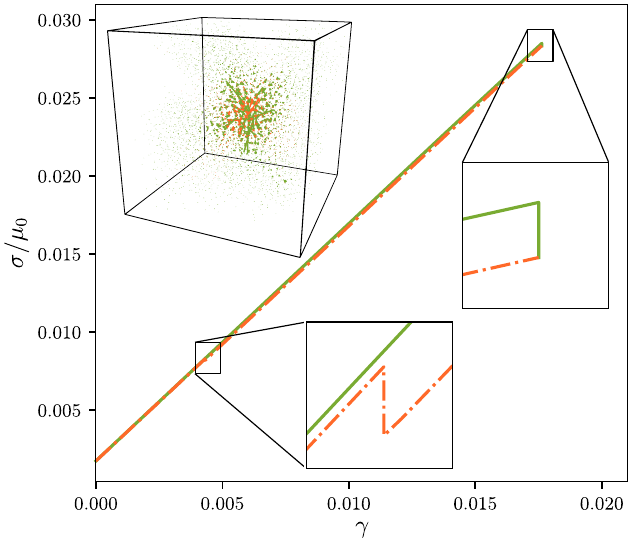}
    \caption{An example of a partially reversible event (structural rearrangement) observed in the analysis whose results are presented in Fig.~4 in the manuscript. This example complements the sketch in Fig.~4a therein, which illustrates a fully reversible and a fully irreversible event, using the same color code and structure. The bottom-right insets are zoom-ins on the stress change induced by the event on the loading and unloading branches. The top-left inset shows the displacements induced by the events, with loading in green and unloading in orange. It is evident that the unloading curve does not overlap with the loading curve after the unloading event, and although both events share the same center, the displacement induced by the unloading event is smaller, both indicating partial reversibility. As noted in the text, such partially reversible events were counted as irreversible in the table in Fig.~4b in the manuscript as they involve dissipation.}
    \label{fig:partially-reversible-event}
\end{figure}

Using this measure, we first deform the system in coarse steps ($\delta\epsilon\={10}^{-4}$) to detect the presence of instability events. Once detected, we repeat the process with very fine strain steps ($\delta\epsilon\={10}^{-7}$) to accurately isolate events. When an isolated event is detected, it is recorded, along with the configuration just before and after the event, for further analysis (e.g., see Fig.~4c in the manuscript and Fig.~\ref{fig:alternative-drop-dists}). The same procedure is applied to an unloading process, which allows us to determine whether an event is reversible/irreversible, see Fig.~4a-b in the manuscript. In relation to Fig.~4a therein, note that intermediate cases of partially reversible rearrangements are observed (deemed irreversible in Fig.~4b in the manuscript because they involve dissipation). An example is presented in Fig.~\ref{fig:partially-reversible-event}.

In Fig.~4c in the manuscript, we quantify stress drops associated with structural rearrangements (instabilities) under both shear and dilation. Therein, the stress drops are made dimensionless using the shear modulus $\mu_0$ and the bulk modulus $K_0$, respectively. In Fig.~\ref{fig:alternative-drop-dists}, we use the very same stress drops dataset, just with two alternative nondimensionalization choices. In Fig.~\ref{fig:alternative-drop-dists}a, we normalize each stress drop by the value of the stress at which it occurs, while in Fig.~\ref{fig:alternative-drop-dists}b, we normalize each stress drop by the value of the relevant state-dependent modulus at which it occurs, see also the legends in Fig.~\ref{fig:alternative-drop-dists}. All three choices support the same physical conclusion: stress drops under dilation are significantly smaller than those under shear.

\vspace{-0.45cm}
\section{M\lowercase{icro-cavities detection algorithm}}
\label{sec:microcavities}

Here, we describe the algorithm developed for micro-cavities detection, which is used in the manuscript. The employed algorithm is adapted from the one discussed in Sect.~II.D of~\cite{debendetti2018mbs}, originally used to identify pores in binary liquids. We first define a `minimal void' as the volume of the smallest particle in the system that can be inserted at a given point inside the system without experiencing any repulsive forces. Once found, a minimal void is `spherically inflated' as long as the no-repulsion criterion is met, defining a void. The concatenated volume of such overlapping voids defines a micro-cavity. To implement this procedure in polydisperse glasses, one needs to develop a sufficiently accurate and efficient algorithm to identify `candidate points' for which the above procedure is applied and to explicitly account for polydispersity.

\begin{figure*}[ht]
    \centering
    \includegraphics[width=0.9\textwidth]{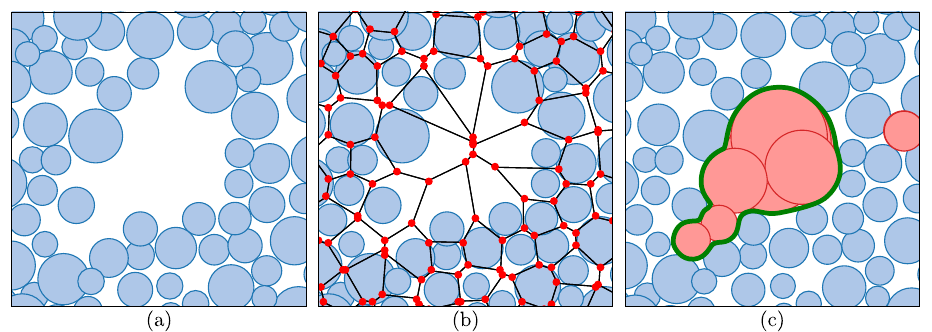}
    \caption{A 2D illustration of the micro-cavities detection algorithm (2D is used here, instead of the actual 3D case analyzed in this work, for visual clarity and simplicity). (a) A small portion of a 2D polydisperse glass, where particles/atoms of various sizes are shown as blue circles. (b) The Voronoi diagram is constructed from the particle positions/centers. Its vertices are marked in red dots and serve as `candidate points' in the detection algorithm, see text for details and explanations. (c) The resulting `inflated spheres' (red circles), i.e., voids, whose radii are determined using Eq.~\eqref{eq:mb-radius}. The cluster of overlapping voids, marked in green, is the detected micro-cavity.}
    \label{fig:cavity-detection-sketch}
\end{figure*}

This is done in two steps. First, we introduce an off-lattice method for identifying candidate points based on the Voronoi diagram of the system~\cite{Voronoi1908}, implemented using the Freud library~\cite{freud2020}. The main merit of this approach is that each Voronoi cell contains exactly one particle and each Voronoi vertex is equidistant and locally farthest-away from the particles on its neighboring cells, making it a natural candidate point for identifying a minimal void (once found, the `sphere inflation' procedure is applied). The off-lattice Voronoi vertices offer a precision that would otherwise require a very dense grid, making the algorithm accurate for large systems, while remaining computationally efficient. It is illustrated in Fig.~\ref{fig:cavity-detection-sketch}.

Second, we adopt a generalized expression for determining the radius of an `inflated sphere' per Voronoi vertex (candidate point), applicable for polydisperse systems, in the form
\begin{align}
    r=\min_{i=1}^m\left[r_i - x_{\mbox{\tiny min}}\lambda_{\min}^i\right] \ .
    \label{eq:mb-radius}
\end{align}
Here, $i$ runs over $m$ particles in the cells incident to the vertex to check for repulsive forces, $r_i$ is the distance between particle $i$ and the vertex, $x_{\mbox{\tiny min}}$ was defined in Sect.~\ref{sec:potentials}, and $\lambda_{\min}^i$ is the pairwise length parameter between particle $i$ and the smallest particle in the polydisperse system. Finally, as already noted above, a micro-cavity is identified as the cluster of overlapping voids.

\begin{figure}[h]
    \centering
    \includegraphics[width=0.9\linewidth]{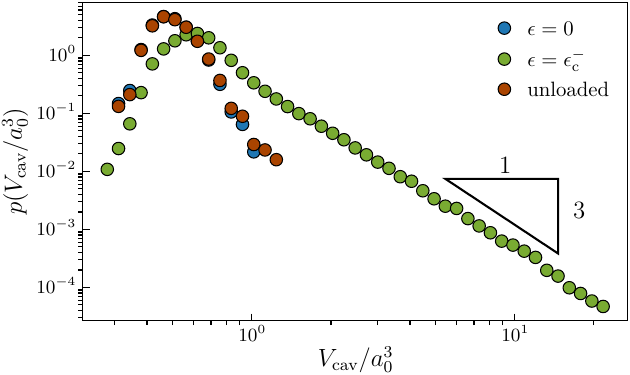}
    \caption{The same as Fig.~6b in the manuscript, but for well-annealed ($T_{\rm p}\!=\!0.35$) LJ glasses. Note that $p(V_{\rm cav}/a_0^3)$ in the unloaded state is very similar to the one in the reference $\epsilon\!=\!0$ state, unlike the corresponding situation for poorly-annealed ($T_{\rm p}\!=\!1.50$) glasses in Fig.~6b in the manuscript, demonstrating the lower degree of glassy disorder in the former. Note, on the other hand, that for the near-cavitation $\epsilon\!=\!\epsilon_{\rm c}^{-}$ state, we observe $p(V_{\rm cav}/a_0^3)\!\sim\!(V_{\rm cav})^{-3}$ at large values, as in Fig.~6b in the manuscript. The degree of universality of this power-law tail should be further tested in future work.}
    \label{fig:cavity-vol-dist-low-Tp}
\end{figure}

The results of applying this algorithm to computer glasses under hydrostatic tension are discussed in the manuscript for poorly-annealed (high $T_{\rm p}$) glasses in Fig.~6 therein. We note that for visualization clarity, we do not show in Fig.~6a in the manuscript all detected micro-cavities (whose minimal volume is that of the smallest particle, by construction), but rather micro-cavities whose volume is larger than the value corresponding to the vertical dashed line in Fig.~6b. The counterpart results to those of Fig.~6b  for well-annealed (low $T_{\rm p}$) glasses are presented in Fig.~\ref{fig:cavity-vol-dist-low-Tp}.

\vspace{-0.45cm}
\section{A \lowercase{first-principles zero-strain nonlinear elastic expansion}}
\label{sec:nonlinear_expansion}
\vspace{-0.45cm}

Here, we briefly discuss the first-principles zero-strain nonlinear elastic expansion used in Fig.~5 in the manuscript, which constitutes one of our major results. Our discussion closely follows~\cite{richard2024connecting} and we refer the reader to it for the full details. The first and second derivatives of the pressure with respect to the dilatational strain parameter $\epsilon$ can be expressed (in $\dbar$ dimensions) as
\begin{equation}
\frac{d P}{d\epsilon}\Bigg|_{\epsilon=0}=-\dbar\,P + \frac{1}{\dbar\,V_0}\frac{d\mathcal{O}_{P}}{d\epsilon} \ ,
\label{eq:first_order}
\end{equation}
and
\begin{equation}
\frac{d^2 P}{d\epsilon^2}\Bigg|_{\epsilon=0}=\dbar^2 P - \frac{2}{V_0}\frac{d\mathcal{O}_{P}}{d\epsilon}+\frac{1}{\dbar\,V_0}\frac{d^2 \mathcal{O}_{P}}{d\epsilon^2} \ ,
\label{eq:second_order}
\end{equation}
respectively, where $\mathcal{O}_P\=-\partial U /\partial \xv \cdot (\calBold{I}\cdot \xv)$, with the identity matrix $\calBold{I}$. The full derivatives $d\mathcal{O}_{P}/d\epsilon$ and $d^2\mathcal{O}_{P}/d\epsilon^2$ include both first-order and second-order nonaffine response contributions, whose explicit expressions can be found in~\cite{richard2024connecting}. It is crucial to note that the derivatives in Eqs.~\eqref{eq:first_order}-\eqref{eq:second_order} are evaluated at $\epsilon\=0$, as highlighted therein, implying that the input to the calculation is just the zero-strain configuration (and of course the interaction potential). The bulk modulus $K_0$ and the second-order dilatational elastic constant $K_2$ of Eq.~(1) in the manuscript are then obtained.

\begin{figure}[!ht]
    \centering
    \includegraphics[width=\linewidth]{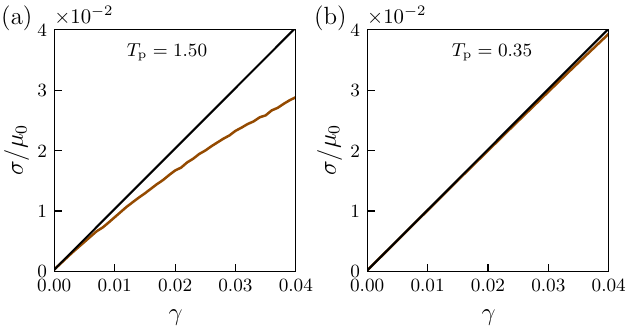}
    \caption{The shear counterparts of the dilation results presented in Fig.~5 in the manuscript, where the zero-strain elastic expansion is considered to first order (black lines, of unity slope), see text for details and discussion.}
    \label{fig:shear_expansion}
\end{figure}

In Fig.~5 in the manuscript we present the results for both poorly- and deeply-annealed dilated glass samples, revealing remarkable agreement between the above-discussed zero-strain nonlinear elastic expansion and the finite $\epsilon$ simulational stress-strain curves nearly up to the large-scale cavitation strain. In Fig.~\ref{fig:shear_expansion}, we present the corresponding results for the very same two glass realizations driven under shear, but up to linear order in the zero-strain expansion, i.e., considering only the shear modulus $\mu_0$. The results reveal qualitative differences compared to the dilation results of Fig.~5 in the manuscript. Specifically, for the poorly-annealed glass sample in Fig.~\ref{fig:shear_expansion}a, the first-order zero-strain nonlinear elastic expansion badly fails due to extensive plasticity (see Fig.~3b in the manuscript), while for the deeply-annealed glass sample in Fig.~\ref{fig:shear_expansion}b, the first-order zero-strain elastic expansion captures well the simulational stress-strain curve due to the predominantly reversible nature of the deformation (see Fig.~3d in the manuscript).

\vspace{-0.4cm}
\section{C\lowercase{u}Z\lowercase{r and} S\lowercase{i}O$_2$ \lowercase{loading-unloading trajectories}}
\label{sec:universality}

\begin{figure*}[h]
    \centering
    \includegraphics[width=0.92\linewidth]{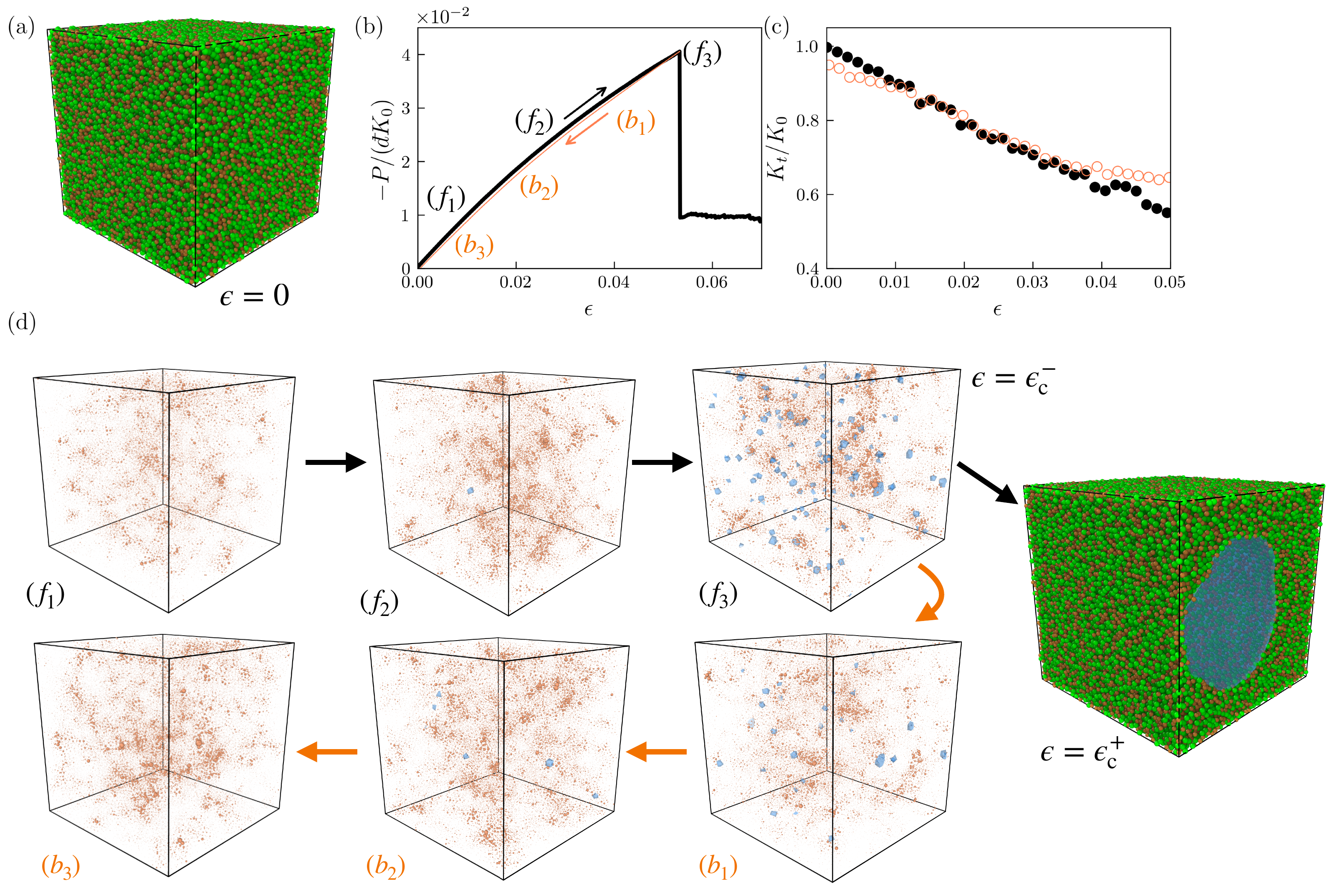}
    \vspace{-0.3cm}
    \caption{(a) The as-cast CuZr glass sample. (b) $-P(\epsilon)/(\dbar K_0)$ as in Fig.~7a in the manuscript, showing the loading branch (in black) and the unloading from $\epsilon_{c}^{-}$ branch (in orange). Various state points are labelled. (c) The tangent bulk modulus $K_{\rm t}(\epsilon)/K_0$ along the loading-unloading path. (d) Snapshots illustrating micro-cavities (in blue) and nonaffine motion (in orange)  at the state points marked in panel (b), and the formation of large scale spherical cavity at $\epsilon_{\rm c}^{+}$. See text for discussion.}
    \vspace{-0.45cm}
    \label{fig:bmg}
\end{figure*}
\begin{figure*}[ht!]
    \centering
    \includegraphics[width=0.92\linewidth]{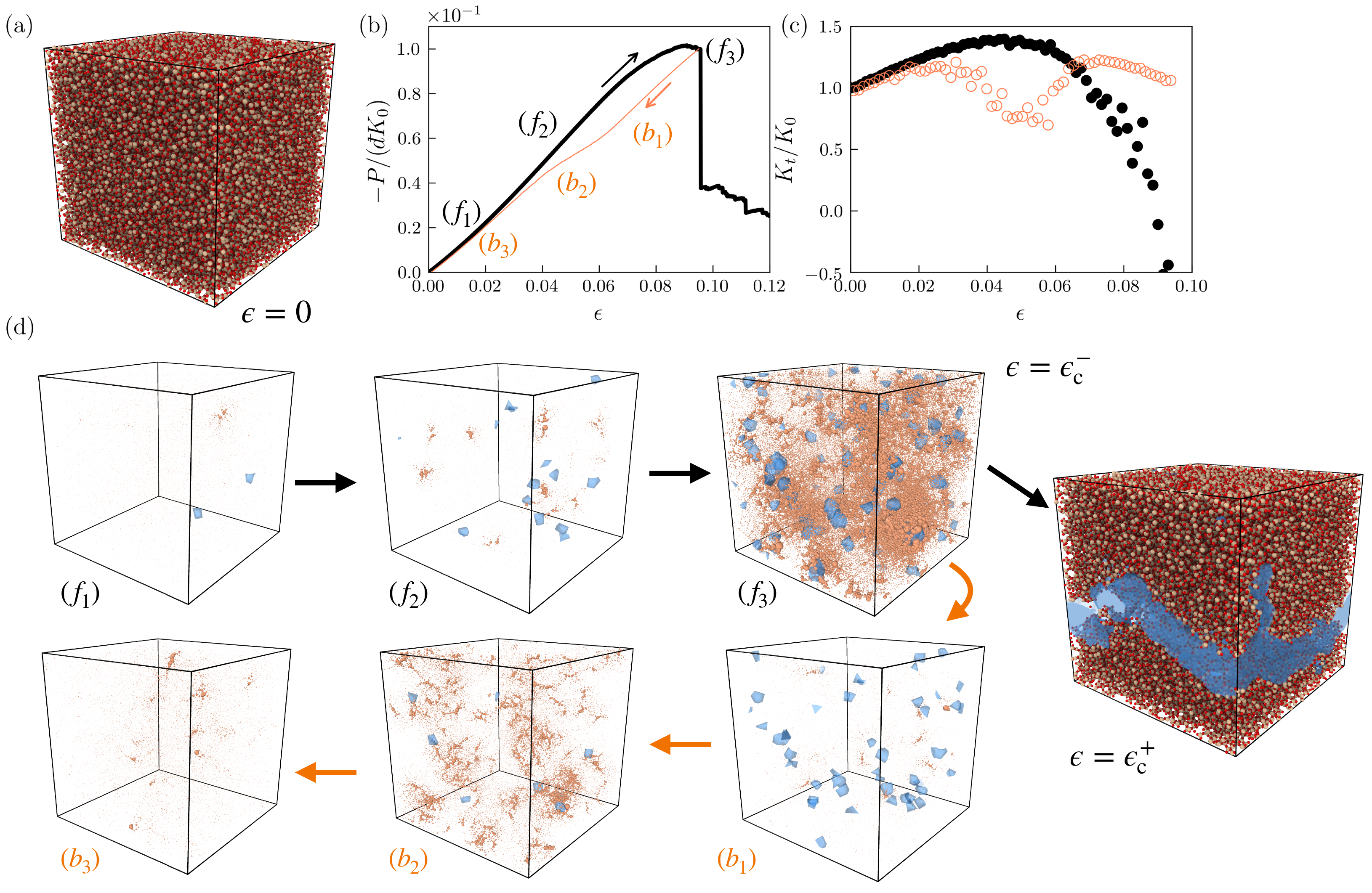}
    \vspace{-0.3cm}
    \caption{The same as Fig.~\ref{fig:bmg}, but for silica glass. See extensive discussion in the text.}
    \vspace{-1cm}
    \label{fig:silica}
\end{figure*}
\begin{figure}[ht!]
    \centering
    \includegraphics[width=\linewidth]{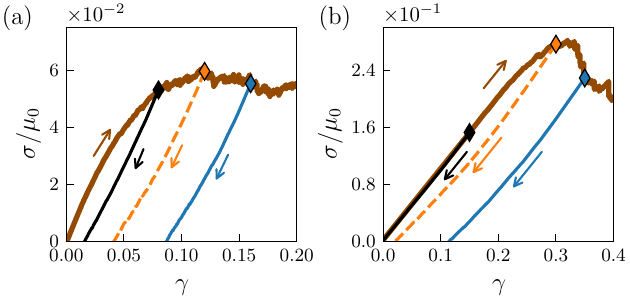}
    \caption{(a) Loading-unloading shear stress-strain curves for (a) CuZr glass and (b) silica glass. See text for discussion.}
    \label{fig:shear_bmg_silica}
\end{figure}

Here, we present additional results in relation to the CuZr metallic glass and SiO$_2$ glass results presented in Fig.~7 in the manuscript. The results in Fig.~\ref{fig:bmg} offer additional insight into the CuZr results of Fig.~7a in the manuscript. In panel (a), we present the as-cast glass at the particle level. In panel (b), we present the loading-unloading stress-strain curve as in Fig.~7a in the manuscript, where we label three points (f$_{i=1-3}$) along the loading branch and three points (b$_{i=1-3}$) along the unloading one. In panel (c), we present the corresponding tangent bulk modulus (along both the loading branch, also shown in the inset of Fig.~7a in the manuscript, and the unloading branch).

In panel (d), we present visual snapshots of the micro-cavities population and atomic-scale nonaffine displacements along the loading-unloading trajectory, using the same state labels as in panel (b), following a similar format to that of Fig.~6a in the manuscript. In this case, we use the Ovito software~\cite{stukowski2010visualization} in a way that: (i) a surface mesh that identifies the micro-cavities is shown in blue, and (ii) particles that exhibit significant nonaffine motion over an incremental strain $\Delta\epsilon\=0.005$ from two consecutive snapshots are shown in orange. Particle radii are equal to the magnitude of the nonaffine displacement.

Overall, the results for the CuZr glass --- generated by a large quench rate leading to a poorly-annealed state --- in Fig.~\ref{fig:bmg} are similar to the LJ results presented in the manuscript, see discussion therein. We note, though, that for the system sizes accessible to us (in term of computational feasibility), no micro-cavities survive unloading back to $\epsilon\=0$ (see snapshot b$_3$). Note also that in the cavitated snapshot, i.e., the $\epsilon\=\epsilon_{\rm c}^{+}$ one, we allow the nearly spherical large-scale cavity to intersect the boundary (instead of centering it as in Fig.~6a in the manuscript, using the periodic boundary conditions) in order to also visualize the particles themselves.

In Fig.~\ref{fig:silica}, we follow the same format of Fig.~\ref{fig:bmg} and present the counterparts of the silica glass results of Fig.~7b in the manuscript. We highlight here some differences compared to the LJ and CuZr results discussed above and in the manuscript. First, we note that the tangent bulk modulus $K_{\rm t}/K_0$ in panel (c) exhibits a non-monotonic behavior, with an initial stiffening of $\sim\!40\%$ over a strain of $\sim\!0.05$, followed by a pronounced softening upon approaching cavitation (in fact, $K_{\rm t}/K_0$ even becomes negative). In the stiffening regime, we observe both nonaffine motion within the silica network and the formation of small micro-cavities (states f$_1$ and f$_2$ in panel (d)). As the system approaches $\epsilon_{\rm c}$, the level of nonaffine motion increases markedly (state f$_3$ in panel (d)) that apparently results in strong softening, culminating in abrupt failure. Second, unlike the LJ and CuZr cases, the spatial mode of failure reveals a strong spherical symmetry-breaking, where instead of an approximately spherical cavity, a rather elongated, crack-like object emerges (see the $\epsilon\=\epsilon_{\rm c}^{+}$ snapshot, where again we allow an intersection with the boundary for visualization purposes, as in Fig.~\ref{fig:bmg}).

Third, the unloading trajectory also reveals rich behaviors. At the onset of unloading, the tangent modulus immediately recovers the as-cast bulk modulus $K_0$ (see open circles in panel (c)), followed by a reduction in the size and number of micro-cavities, with only minute nonaffine rearrangements (state b$_1$ in panel (d)). Subsequently, a significant increase in nonaffine motion is observed (state b$_2$), accompanied by an apparent softening of the tangent modulus (panel (c)), upon which the unloading curve approaches the loading curve (panel (b)). Finally, the micro-cavities disappear, the amplitude of nonaffine displacements progressively decreases (state b$_3$) and the unloading tangent modulus follows the loading one in reverse order (panel (c)). These behaviors call for additional investigation.

For completeness, we present in Fig.~\ref{fig:shear_bmg_silica} the loading-unloading strain-strain curves for the CuZr and silica glasses under shear. As expected, the CuZr behavior in Fig.~\ref{fig:shear_bmg_silica}a is very similar to the poorly-annealed LJ glass behavior in Fig.~3b in the manuscript, both in revealing a continuous yielding transition upon loading and a predominantly elastic unloading, accompanied by a residual plastic strain. For the silica glass, see Fig.~\ref{fig:shear_bmg_silica}b, we observe a stress overshoot along the loading path, which corresponds to the formation of a shear-band. For small strains, the unloading path (black curve) follows a predominantly reversible trajectory with only minute plasticity. As the strain approaches the yield strain, which is smaller than the shear-banding strain, increasing irreversibility is observed (orange path), indeed indicating that significant plastic deformation occurs prior to shear-band formation. Finally, unloading the sample right after shear-banding failure (blue curve) results in a residual plastic strain of order ${\cal O}(0.1)$.


%

\end{document}